\documentstyle[12pt]{article}

\newcommand{\beq}{\begin{equation}}
\newcommand{\eeq}{\end{equation}}

\begin{document}

\rightline{{\bf  US--FT/33--96}}
\rightline{{\bf  hep--ph/9607239}}
\rightline{{\bf  July 1996}}

\begin{center}
\vskip 0.75cm

{\Huge {\bf Percolation approach to Quark Gluon Plasma}}\\
\vskip 0.2cm
{\Huge {\bf and $J/\psi$ suppression}}\\
\vskip 0.8cm

N. Armesto, M. A. Braun$^{*}$, E. G. Ferreiro and C. Pajares\\ 
{\sl Departamento de 
F\'{\i}sica de  Part\'{\i}culas,
Universidade de Santiago de
Compostela,\\ 15706--Santiago de Compostela, Spain}

\vskip 1.0truecm
{\Large {\bf Abstract}}
\end{center}
\begin{quotation}

It is shown that the critical threshold for percolation of the overlapping 
strings exchanged in heavy ion collisions can naturally explain the sharp 
strong suppresion of 
$J/\psi$ shown by the experimental data on central Pb--Pb collisions, 
which does not occur in central O--U and S--U collisions.

\end{quotation}
\vspace{8.5cm}

\noindent 
$^{*}$Permanent address: Department of High Energy
Physics, University of St. Petersburg, 198904 St. Petersburg, 
Russia.

\newpage

The NA50 collaboration (\cite{Gun96}) has reported a strong 
suppression of $J/\psi$ production in Pb--Pb central 
collisions at 158 AGeV/c per nucleon. The suppression is much stronger 
than the expected one due to $J/\psi$ absorption corresponding to a 
cross section of 6.3 mb, by which the 
NA38 data for O--U and S--U central collisions 
(\cite{Huf92,Cap88}) and the hadron--nucleus data
can be explained. The NA50 data show a clear 
deviation from the previous situation
(\cite{Bag89}).
The $J/\psi$ suppression in peripheral Pb--Pb collisions is similar to the one 
corresponding to central S--U collisions, but a sharp enhancement 
occurs as the centrality of the Pb--Pb collisions increases.

In this paper we draw attention to the fact that 
the continuum percolation of colour 
strings can naturally describe the sharp
difference in the 
$J/\psi$ suppression at present energies between O--U, S--U and peripheral 
Pb--Pb collisions on the one side and central Pb--Pb
collisions on the other side.
Predictions for RHIC and LHC energies are given.

The continuum percolation of colour strings takes place 
when the density of strings
rises above a threshold, which can be 
calculated on geometrical grounds. In this
picture, the region where several strings fuse 
can be considered a droplet of a non--thermalized
Quark Gluon Plasma, in which the $J/\psi$ is suppressed as predicted by
Matsui and Satz 
(\cite{Mat86}). Percolation means that 
these droplets overlap and the Quark Gluon Plasma
domain becomes comparable to the nuclear size.

In many models of hadronic collisions
(\cite{And87}--\cite{Kai82}), colour
strings are exchanged between 
projectile and target. The number of strings grows with the energy
and with the number of nucleons of the participant nuclei. When 
the density of strings becomes high the string colour fields begin 
to overlap and eventually individual strings may fuse
(\cite{Paj90}--\cite{Cap96}), forming a new string
which has a higher colour charge at its ends, corresponding to the summation of 
the colour charges located at the ends of the original strings. The 
new strings break into hadrons according to their higher colour. As a 
result, heavy
flavour is produced more efficiently and there 
is a reduction of the total multiplicity (\cite{Ame93}).
Also, as the energy--momenta of the original strings 
are summed to obtain the energy--momentum of 
the resulting string, the fragmentation of the latter can produce 
some particles outside the kinematical limits of nucleon--nucleon 
collisions if the original strings 
come from different nucleons
(\cite{Ele96,And91}). The fusion of strings 
has been incorporated in several Monte Carlo codes. In particular, in the 
Quark Gluon String Model (QGSM)
it is assumed that strings fuse 
when their transverse positions come within a 
certain interaction area $a$
(\cite{Ame93}). The value of $a$ is determined to reproduce
$\overline{\Lambda}$ rapidity distributions in S--S and S--Ag central 
collisions at $p_{lab}$= 200 GeV/c and in Pb--Pb central collisions at 
$p_{lab}$= 158 GeV/c. 

Cascade reactions like 
$\pi^+$$\overline{p}$$\rightarrow$$K^0$$\overline{\Lambda}$, 
$\pi^0$$\overline{n}$$\rightarrow$$K^0$$\overline{\Lambda}$,
$\pi^+$$\overline{\Lambda}$$\rightarrow$$K^+$$\overline{p}$ and 
$p$$\overline{\Lambda}$$\rightarrow$$\pi^+$$\overline{K^0}$
also contribute to the $\overline{\Lambda}$ rapidity 
distribution but their effects are smaller than the ones 
due to string fusion,
generating uncertainties in the value of $a$ of
 around $10\%$. From the value
of $a$, the radius $r$ of the transverse dimension of the
string can be obtained, $a=2\pi r^2$ (\cite{Arm95}).
In our code only fusion of two strings is considered, so the obtained
$r$--value, $r$=0.36 fm, is an effective one, somewhat 
larger than the real transverse
radius of the string. 
Denoting by $N_j$ the number of strings which fuse into $j$--fold strings and
$N_2^{'}$ and $r_{eff}$ the number of all fused strings and the effective
transverse size of the string, respectively, we will have
\beq
2 N_2^{'} \pi r_{eff}^2 = \sum_{j=2} N_j j \pi r^2,
\eeq
\beq
N_2^{'} = \sum_{j=2} N_j.
\eeq
The upper limit of the sum in (1) is determined by the constraint (2). 
The values of $N_2^{'}$ and $r_{eff}^2$ were
fixed 
in our calculation by 
comparing the results of the string fusion model with the experimental
data on $\overline{\Lambda}$ production in central S--S collisions at 
$\sqrt s$
=19.4 AGeV. 
Computing $N_j$ in our Monte Carlo code
 we obtain from (1) the value $r$=0.2 fm
both for Pb--Pb and S--Ag collisions.
 
In nucleus--nucleus collisions many strings are exchanged. In impact
parameter space these strings are seen as circles inside the total 
collision area.
As the number of strings increases, 
more
strings overlap. Several fused strings can be 
considered as a domain of a non--thermalized Quark
Gluon Plasma. Following the arguments of Matsui and Satz (\cite{Mat86})
the $J/\psi$ can not be formed 
inside this domain.
Also, the $J/\psi$ will be
destroyed by interaction with these fused strings. Above a critical density of
strings percolation occurs, so that paths 
of overlapping circles are formed through the whole
collision area. Along
these paths the medium behaves like a colour conductor. The percolation gives
rise to the formation of Quark Gluon Plasma on a nuclear scale. 
The phenomenon of continuum percolation is well known
(\cite{Isi92}). 
It explains 
hopping conduction in
doped semiconductors and other important physical processes
(\cite{Shk84}). The percolation
threshold $\eta_c$ is related to the critical density of circles $n_c$ by the
expression

\beq
\eta_c=\pi r^2 n_c.
\eeq
$\eta_c$ has been computed using Monte Carlo simulation, direct--connectedness
expansion and other different methods. All the results are in the range
$\eta_c=1.12-1.175$
(\cite{Pik84}--\cite{Dom60}). 
Taking the above mentioned value of $r$, these values
imply 

\beq
n_c=8.9-9.3 \ \  {\rm strings/fm}^2.
\eeq

One may introduce a hard core to model a repulsive interaction between the
circles, or to substitute 
circles by squares. The percolation threshold $\eta_c$
is only slightly reduced in these cases. This enhances the confidence in its
value and the application to our case where we do not know the dynamics of the
interaction among strings.

In Table 1 the number of strings exchanged for central p--p, S--S,
S--U and Pb--Pb collisions 
is shown together with their densities. It is seen that at SPS
energies only the density reached in central Pb--Pb collisions is above the
critical density. In Pb--Pb minimun bias collisions the average number of
strings at SPS energies is 227, very similar to the value for central S--U
collisions, so the density is lower than the
critical one. 

The $J/\psi$ suppression experimentally observed follows the same pattern. The
strong suppression is only observed in central Pb--Pb collisions. According to
Table 1, a strong $J/\psi$ suppression is also expected in S--U collisions at
RHIC energies and in S--S and S--U collisions at LHC energies.

Recently
(\cite{Bla96})
 it has been assumed that the produced $J/\psi$ 
is completely 
destroyed whenever the energy density exceeds a certain
value and this energy density is taken 
proportional to the density of participants.
The critical value is chosen to lie between the density of participants of
central S--U collisions and Pb--Pb collisions. With this choice a good
description of the experimental data is obtained. In our model the density
of strings is proportional to the number of collisions, and we obtain similar
quantitative results. However, in our approach the critical value is naturally
explained on geometrical grounds. 

In Fig. 1 the distribution of strings fusing into sets
of 
a given number of 
fused strings is shown for central S--U collisions at $\sqrt s$=19.4 AGeV
and  $\sqrt s$=200 AGeV 
and also for central Pb--Pb collisions at $\sqrt s$=19.4 AGeV. 
The first case is below and the second above the
percolation threshold. It is seen that above the percolation threshold 
we can obtain many sets with a very high
number of fused strings. 

Refering to $\psi^{'}$ suppression the experimental data reveal the following
features 
(\cite{Gun96},
\cite{Lou95}--\cite{Ald91}):

\noindent 1) The ratio $\psi^{'}/\psi$ is constant in p--A collisions.

\noindent 2) $\psi^{'}/\psi$ decreases with centrality in S--U collisions.
The decrease seems to stop at high centrality.

\noindent 3) $\psi^{'}/\psi$ is almost the same in central Pb--Pb and S--U
collisions.

The first two features of experimental data can be explained by 
absorption and
interaction with comovers
(\cite{Gav90}--\cite{Won96}). Taking equal absorption cross section 
$\sigma(J/\psi)=\sigma(\psi^{'})\sim 4.2$ mb, 
the hadron--nucleus behaviour can be
explained since no interaction with the produced particles
(\cite{Won96}) is assumed.
In nucleus--nucleus collisions low energy interactions
of $J/\psi$ and $\psi^{'}$ with the hadrons produced in the collision
break both the $\psi^{'}$ and the $J/\psi$ but the $\psi^{'}$ 
cross section
at low energy is much larger than that of $J/\psi$ (the threshold
for breaking the $\psi^{'}$ 
is only 52 MeV and the one for $J/\psi$
is 640 MeV). This difference in the cross sections may be 
responsible for the different behaviour of $\psi^{'}$ and
$J/\psi$ suppression 
in central S--U collisions. In our picture this behaviour can be
explained 
by noting that in S--U central collisions 
the average distance between strings is the order of 0.4 fm, 
larger than the size of $J/\psi$ (0.2 fm) but less than the size of the
$\psi^{'}$. Therefore one expects that $\psi^{'}$ interacts with the
strings or with the particles produced by the strings 
with greater probability than
$J/\psi$. For central Pb--Pb collisions, with the density above the critical
percolation threshold, no additional suppression of $\psi^{'}$ 
relative to
$J/\psi$ is expected, in agreement with the data. 

Also it is possible that the percolation process takes place among the
produced resonances and particles instead of strings
(\cite{Aic93}). The two cases can be
distinguished by studying the behaviour of long range correlations and
measuring forward--backward correlations (\cite{Ele94}).

The percolation of strings can be considered as a smooth way to Quark Gluon
Plasma. Around percolation threshold, strong fluctuations in the number
of strings with a given colour should appear. This will produce
large fluctuations in a number of different observables,
like strangeness, in
an event by event analysis. 
Also 
a large number of $\Omega^-$ (confirmed by the experimental data 
(\cite{WA97})) and a copious production of hadronic particles with
$|x_F|$ much larger than 1, outside the kinematical nucleon--nucleon limits, 
may serve as clear
signatures. The latter would also 
distinguish our picture from the percolation of
resonances and particles. 

In conclusion we thank A. Capella
for useful comments and discussions and the 
Comisi\'on Interministerial de Ciencia y Tecnolog\'{\i}a (CICYT)  
of Spain for financial support under contract AEN96-1673. 
Also M. A. Braun thanks
IBERDROLA and E. G. Ferreiro 
the Xunta de Galicia for financial support.

\newpage
\noindent{\Large {\bf Table captions}}

\vskip 0.5cm

\noindent{\bf Table 1.} 
Number of strings (upper numbers) and their densities (fm$^{-2}$) (lower
numbers) in central p--p, S--S, S--U and Pb--Pb collisions at SPS, RHIC and LHC
energies.

\newpage

\newpage
\noindent{\Large {\bf Figure captions}}

\vskip 0.5cm

\noindent{\bf Figure 1.}
Percentage of the total number of strings exchanged in the collision which goes
into sets of a given number of fused strings, for central S--U collisions at
$\sqrt s$=19.4 AGeV (dashed line)
and  $\sqrt s$=200 AGeV (solid line)
and for central Pb--Pb collisions at $\sqrt s$=19.4 AGeV (dotted line).
The number 10 in the horizontal axis indicates sets of 10 or more strings.

\newpage

\begin{center}
{\bf Table 1}
\vskip 1cm
\begin{tabular}{ccccc} \hline
\hline
$\sqrt s \ \ \rm (AGeV)$ & & Collision \\ \hline
& $p-p$ & $S-S$ & $S-U$ & $Pb-Pb$  \\ \hline
19.4 & 4.2 & 123 & 268 & 1145 \\ 
& 1.3 & 3.5 & 7.6 & 9.5 \\
\hline
200 & 7.2 & 215 & 382 & 1703 \\
& 1.6 & 6.1 & 10.9 & 14.4 \\
\hline
5500 & 13.1 & 380 & 645 & 3071 \\
& 2.0 & 10.9 & 18.3 & 25.6 \\
\hline
\end{tabular}
\end{center}

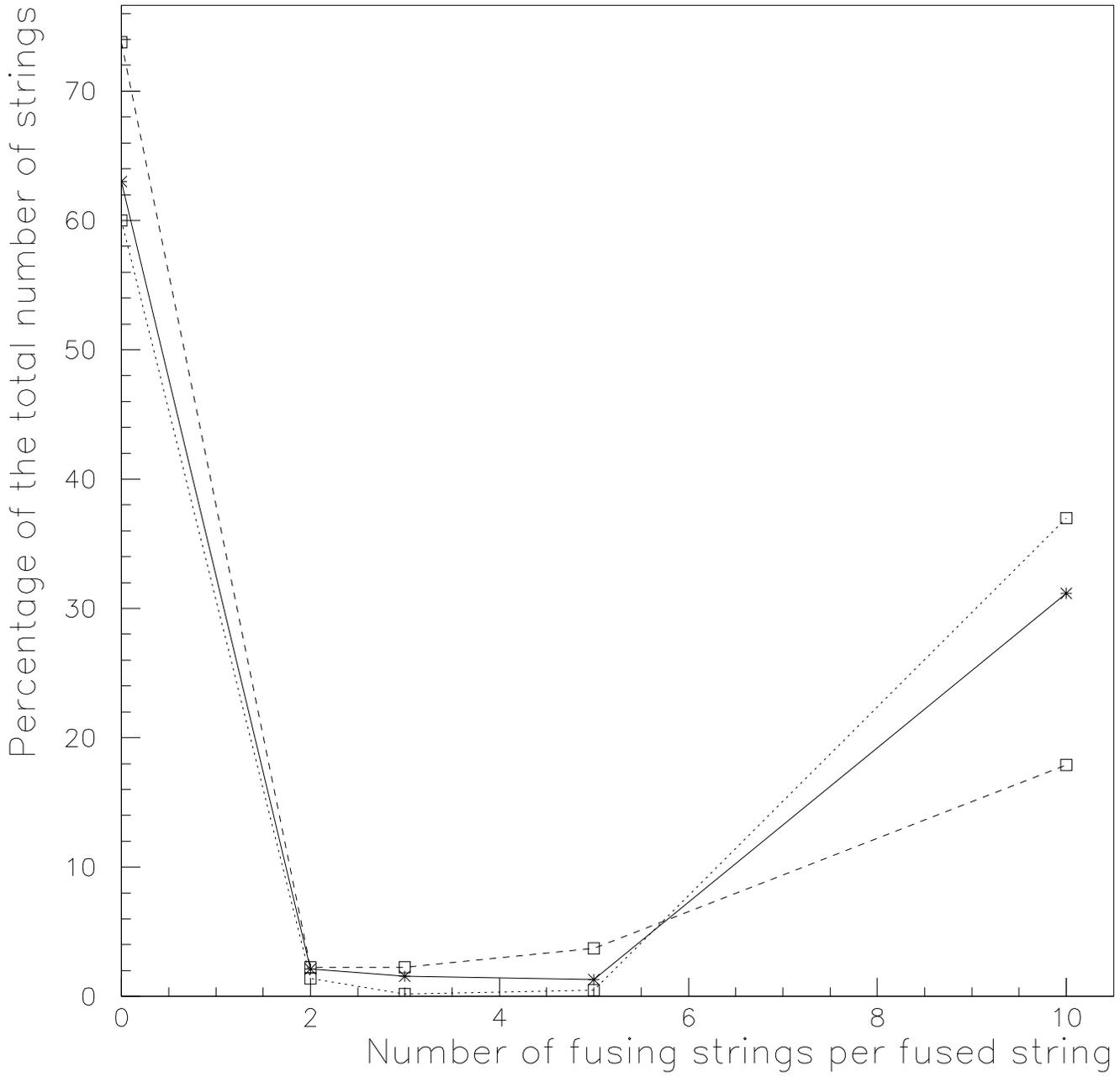
\end{document}